\documentclass[11pt,a4paper,showpacs,superscriptaddress]{revtex4}
\usepackage[utf8]{inputenc}
\usepackage{amsmath}
\usepackage{graphicx}
\usepackage{amssymb, axodraw4j}

\makeatletter
\@ifundefined{textcolor}{}
{%
 \definecolor{BLACK}{gray}{0}
 \definecolor{WHITE}{gray}{1}
 \definecolor{RED}{rgb}{1,0,0}
 \definecolor{GREEN}{rgb}{0,1,0}
 \definecolor{BLUE}{rgb}{0,0,1}
 \definecolor{CYAN}{cmyk}{1,0,0,0}
 \definecolor{MAGENTA}{cmyk}{0,1,0,0}
 \definecolor{YELLOW}{cmyk}{0,0,1,0}
 }



\usepackage{hyperref}

\makeatother

\begin{document}

\title{Induced Chern-Simons Like Action by Lorentz Symmetry Breaking in (3+1)D QED}

\author{D. M. Oliveira}

\email{denny@fma.if.usp.br}

\affiliation{Instituto de F\'isica, Universidade de S\~ao Paulo, Caixa Postal 66318, 05315-970, S\~ao Paulo - SP, Brazil}

\begin{abstract}
In this paper, the induced Chern-Simons like action for a
system of fermions interacting with a gauge field in (3 +1) dimensions
is calculated in the presence of a background field that breaks the Lorentz and CPT
symmetries. The main result of this work corresponds to the obtation
of a usual Chern-Simons-type action in four-dimensional spacetime
resulting from the addition to the conventional QED Lagrangian of a term with a background field that
breaks the Lorentz symmetry of the Lagrangian gauge theory. It is
pointed out that the proportionality constant in that term depends
heavily on the regularization method used in the computations. As the articles on this subject in the literature are very
difficult to read, here calculations are performed clearly and objectively.
\end{abstract}

\maketitle

\section{\label{sec:I}Introduction}

The CPT theorem \cite{Schwinger1951} is a well-known property of the
Standard Model (SM) of elementary particle physics. The Lorentz
symmetry, together with the CPT theorem, is an important mechanism for
understanding the phenomena that occur within the quantum
theory. However, in recent years, the Lorentz symmetry has been broken
in an attempt to incorporate the theory of General Relativity to the
SM \cite{Mariz2004,Gomes2008b}. An initial idea for a possible theory of Lorentz
symmetry breaking appeared in an work of Kosteleck\'y and Samuel
\cite{Kostelecky1989}. In that study, the authors argued that such
violation can be extended to the SM. Then, in the second half of 1990,
an extension of the regular SM, the Standard Model Extension (SME),
came about \cite{Kostelecky1989,Colladay1997,Colladay2001}. \par

The SME is a theory that has all of the usual SM properties - such as
the gauge structure $SU(3)\times SU(2)\times U(1)$ and
renormalizability - and the extension that allows for violations of
Lorentz and CPT symmetries. This theory then provides a quantitative
description of the violations of Lorentz and CPT symmetries,
controlled by coefficients whose values are determined by especific
experiments. A striking feature of this theory, as is known from the
literature, is that the breaking of the CPT symmetry implies Lorentz symmetry
breaking \cite{Greenberg2002}. This fact means that any observable
violation of CPT symmetry is described by the SME \cite{Bluhm2001}. In
the reference \cite{Colladay1997}, a theoretical basis to perform
perturbative calculations in this theory, via fermion sector, can be
found. \par

In the early 1980's, S. Deser \cite{Deser1982} wrote the Maxwell
electromagnetic theory in a planar variant in (2 +1)$D$, which
preserves Lorentz invariance and gauge transformations. This model,
called the theory of Maxwell-Chern-Simons (MCS) \cite{Oliveira2011},
has applicability to planar condensed matter phenomena, with great
emphasis in the literature to superconductors and the fractional
quantum Hall effect \cite{Dunne1999}. However, in a 1990 work,
Carroll, Field and Jackiw realized in a pioneering work
\cite{Carroll1990} that it is possible to formulate a similar theory in (3 +1) dimensions, by adding the action
\begin{equation}
 S_{CS}^{(3+1)D}=\frac{1}{2}\int d^4x\,\varepsilon^{\mu\nu\rho\sigma}\eta_\mu A_\nu \partial_\rho A_\sigma
\end{equation}
to the conventional Maxwell action. This term is known in the
literature as the Carroll-Field-Jackiw term \cite{Jackiw1999}. This
term is CPT-odd. Although gauge transformations may be preserved, the
Lorentz symmetry is violated because it is necessary to engage the end
type Chern-Simons (CS) $\eta_\mu$, a constant four-vector, which
produces an anisotropy of spacetime \cite{Dunne1999}. \par

In this paper, radiative corrections in the approximation of a loop
are calculated from the axial coupling of fermions with a gauge field
in the presence of a Lorentz symmetry breaking term. This coupling
generates an induction of a term similar to CS \cite{Jackiw1999}, as
in equation (1), in the action of Quantum Electrodynamics (QED). The
induction is ambiguous, since the proportionality between the fields
of matter and radiation depends exclusively on the regularization
scheme adopted (such schemes can be: dimensional regularization,
Pauli-Villars \cite{Altschul2004} regularization, the method of
cut-off \cite{Brito2008} and the Schwinger propper time method
\cite{Mariz2008}). Thus, the induced term by the method of dimensional
regularization is computed. The calculations are performed in a clear way, since the articles on the subject in the literature are difficult to understand and the calculations are very tedious. This is the main goal of this work.

\section{\label{sec:II}Fermion propagator expansion and the breaking of Lorentz symmetry}

The fermion propagator dependent of Lorentz symmetry breaking \cite{Colladay1997} is given by
\begin{equation}
S_b(p)=\frac{i}{\not\!p-m-\not\!b\gamma_5}\,.
\end{equation}

When inverted, the above propagator is represented by \cite{Perez-Victoria1999}
\begin{equation}
 S_{b}(p)=\frac{i(\not\!p-\not\!b\gamma_5+m)\{ p^2-b^2-m^2+[\not\! p,\not
b]\gamma_5 \}}{(p^2-b^2-m^2)^2-4(p\cdot b)^2+4p^2b^2}\,.
\end{equation}

This propagator has a complicated structure and makes it more tedious
perturbative calculations (exact perturbative calculations with this
propagator can be found in the reference
\cite{Perez-Victoria1999}). This propagator will not be used here, but
an alternative method will be drawn. As the four-vector $b_\mu$, which
signals the Lorentz symmetry breaking in the new theory, is very small
compared to the electron mass, the correction made in the propagator
can be treated in a perturbative way. So the following expansion in a
sum of terms of an infinite geometric progression is applied
\begin{equation}
 S_b(p)=\frac{i}{\not\!p-m}+\frac{i}{\not\!p-m}(-i\not\!b\gamma_5)\frac{i}{\not\!p-m}+\frac{i}{\not\!p-m}(-i\not\!b\gamma_5)\frac{i}{\not\!p-m}(-i\not\!b\gamma_5)\frac{i}{\not\!p-m}+\cdots
\end{equation}

Thus, with each $\times$ representing each insertion $-\not\!b\gamma_5$ in the propagator, its graph is given by
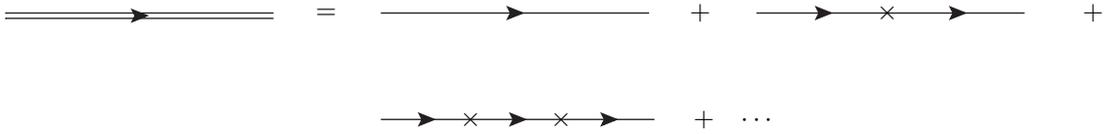
\begin{figure}[h]
\centering
\hspace*{-9cm}\begin{picture}(150,50)
 
\ArrowLine(49,19)(51,19)
\Line(0,18)(100,18)\Text(120,20)[]{=}
\Line(0,20)(100,20)

\ArrowLine(140,20)(240,20)\Text(260,20)[]{+}

\ArrowLine(280,20)(330,20)\Text(330,20)[]{$\times$} 
\ArrowLine(330,20)(380,20)\Text(407,20)[]{+}

\ArrowLine(140,-20)(174,-20)\Text(174,-20)[]{$\times$} 
\ArrowLine(174,-20)(208,-20)\Text(208,-20)[]{$\times$}
\ArrowLine(208,-20)(242,-20)\Text(273,-20)[]{+\hspace{1eM}$\cdots$}
\end{picture}
\vspace{1cm}
\caption{Feynman diagrams of the expansion of the propagator (4). Each
mark $\times$ represent an insertion of the term $-\!\not\!b\gamma_5$ in the propagator.}
\end{figure}
\section{\label{sec:III}Calculation of the induced  term}

The radiative corrections to the action of the usual QED will be
calculated by considering a system of fermions coupled to a gauge
field $A_\mu$ formulated in the spacetime in (3 +1) dimensions. Using
the disturbance $\not\!b\gamma_5$, the Lagrangian of this model is
given by \cite{Oliveira2010}
\begin{equation}
 {\cal L}=\bar{\psi}(i\!{\not\!\partial}-\not\!b\gamma_5-e\!{\not\!\!A} -m)\psi\,.
\end{equation}

To calculate the induced term, the path integrals formalism will be
used \cite{Feynman1965}. The effective action for this model depends
on the term of the Lorentz symmetry breaking
$\bar{\psi}\not\!b\gamma_5\psi$, which, in a one loop approximation, is defined as follows \cite{Gomes2002}
\begin{equation}
 e^{iS_{eff}[b,m]}=N\int D\bar{\psi}D\psi\,exp\left[i\int d^4x\,\bar{\psi}(i\!{\not\!\partial}-e\not\!\!A-\not\!b\gamma_5-m)\psi\right]\,,
\end{equation}
where $N$ is a normalization constant.
\smallskip

With the use of Grassmann variables, integrating over the fermions
fields, it yields
\begin{equation}
 e^{iS_{eff}[b,m]}=N\,det(i\!{\not\!\partial}-e\not\!\!A-\not\!b\gamma_5-m)\,,
\end{equation}
or, in an alternative way,
\begin{equation}
 S_{eff}[b,m]=-iTr\ln[i\!{\not\!\partial}-e\not\!\!A-\not\!b\gamma_5-m]\,.
\end{equation}

Where $A$ and $B$ are two matrices that do not commute, the following
identity is obtained:
\begin{equation}
 \ln(B-A)=\ln B-\sum\limits_n \frac{1}{n}\left[\frac{1}{B}A\right]^n\,.
\end{equation}

Identifying $A=e\not\!\!A$ and
$B=i\!{\not\!\partial}-\not\!b\gamma_5-m$ in the expression (9), it
follows that
\begin{eqnarray}
 S_{eff}[b,m]=-iTr\ln[i\!{\not\!\partial}-\not\!b\gamma_5-m]+iTr\sum\limits_{n=1}^{\infty}\frac{1}{n}\left[ \frac{1}{i\!{\not\!\partial}-\not\!b\gamma_5-m}e\not\!\!A  \right]^n\,.
\end{eqnarray}

The first term of the above expansion corresponds to a constant term added to the action and therefore does not matter for our purposes, since it does not depend on the gauge field $A_\mu$. The contributions come from terms for $n>1$. Thus, for $n = 1$ in the expansion above, 
\begin{equation}
 S_{eff}^{(1)}=ie\,Tr\int d^4x\int\frac{d^4p}{(2\pi)^4}\frac{1}{\not\!p-\not\!b\gamma_5-m}\not\!\!A\,.
\end{equation}

The contributions of (11) give rise to terms like tadpoles, which are
linear in $A_\mu$ and are divergent in the ultraviolet. Because these
terms do not contribute to the induction of the CS like term, they
will be disregarded in the calculations below. However, we illustrate
their graphics to the first order in gauge field in Figure 2.

\begin{figure}
\hspace{-1.1cm} \begin{picture}(500,250)(0,0)
  \Photon(0,200)(80,200){3}{4} \BCirc(100,200){20} \BCirc(100,200){18.3} \ArrowLine(119,201)(119,199) \Text(140,200)[]{=}

\Photon(160,200)(240,200){3}{4} \BCirc(260,200){20} \ArrowLine(280,201)(280,199) \Text(300,200)[]{+}

\Photon(320,200)(400,200){3}{4} \ArrowArcn(420,200)(20,180,0) \ArrowArcn(420,200)(20,360,180)\Text(441,200)[]{$\times$}
\Text(460,200)[]{+}

\Photon(60,100)(140,100){3}{4}\ArrowArcn(160,100)(20,180,90) \ArrowArcn(160,100)(20,90,270) \ArrowArcn(160,100)(20,270,180)
\Text(161,120)[]{$\times$}  \Text(161,80)[]{$\times$} \Text(216,100)[]{+}

\Photon(250,100)(330,100){3}{4} \ArrowArcn(350,100)(20,180,90)   \ArrowArcn(350,100)(20,90,0)   \ArrowArcn(350,100)(20,0,270)  \ArrowArcn(350,100)(20,270,180)  
 
\Text(350,120)[]{$\times$} \Text(371,100)[]{$\times$}
\Text(350,80)[]{$\times$}   \Text(420,100)[]{+\hspace{3eM}$\cdots$} 
\end{picture}
\vspace*{-3.15cm}
\caption{Tadpole representations of the Feynman diagrams to the first
  order of gauge field.}
\end{figure}
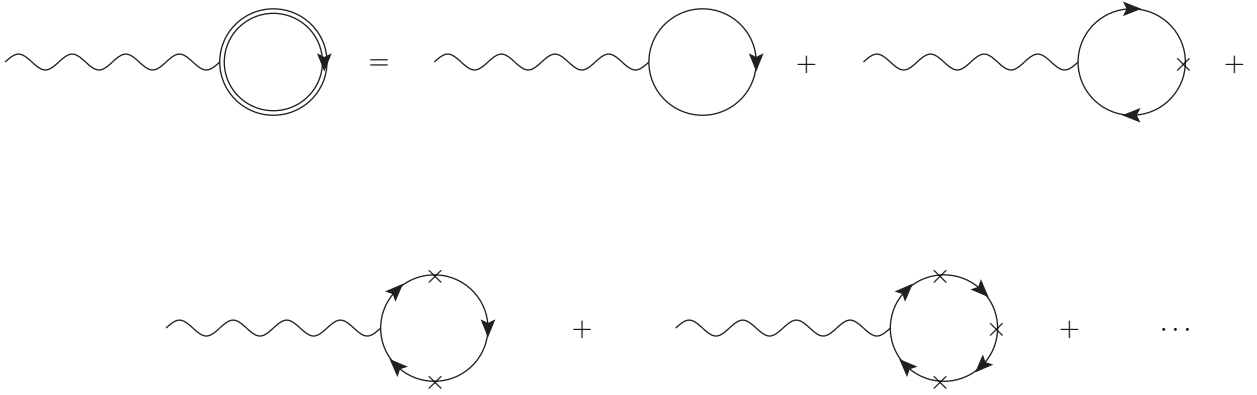

The second order contributions provide
\begin{equation}
S_{eff}^{(2)}= -\frac{ie^2}{2}Tr[S_b(p)\not\!\!AS_b(p)\not\!\!A]\,.
\end{equation}

The calculation of the trace shown above for the effective action is
similar to the calculation of the trace of an operator ${\cal O}$
which depends on the Dirac matrices and internal indices of the Lie
group. So, the total trace $Tr$ represented in (12), in (3 +1) dimensions, is defined by:

\begin{equation}
 Tr{\cal O}\,\dot{=}tr_D\int d^4x\langle x|{\cal O}|x'\rangle\bigg|_{x=x'}\,,
\end{equation}
where the symbol $tr_D$ indicates that the trace will be calculated for the Dirac matrices in standard representation.
\smallskip

The result of this calculation is given by
\begin{equation}
 S_{eff}^{(2)}=\frac{ie^2}{2}tr_D \!\int\! d^4x \!\int\!\frac{d^4p}{(2\pi)^4}\frac{1}{\not\!p-\not\!b\gamma_5-m}\not\!\!A(x)\frac{1}{\not\!p-i\!{\not\!\partial}-\not\!b\gamma_5-m}\not\!\!A(x)\,.
\end{equation}

Using the change of variables $p - q \to k$, it yields
\begin{equation}
 S_{ef}^{(2)}=\frac{ie^2}{2}tr_D \!\int\! d^4x \!\int\!\frac{d^4p}{(2\pi)^4}S_b(p)\not\!\!AS_b(p-i\!{\not\partial})\not\!\!A\,.
\end{equation}

\begin{figure}
\hspace{-0.7cm} \begin{picture}(500,340)(0,0)

\Photon(0,300)(80,300){3}{5} \BCirc(105,300){25} \BCirc(105,300){23} \ArrowLine(104,324)(106,324) \ArrowLine(106,276.2)(104,276.2)
\Photon(130,300)(210,300){3}{5} \Text(225,300)[]{=}

\Photon(240,300)(320,300){3}{5}\ArrowArcn(345,300)(25,180,0) \ArrowArcn(345,300)(25,360,180)\Photon(370,300)(450,300){3}{5}
\Text(470,300)[]{+}

\Photon(0,200)(80,200){3}{5} \ArrowArcn(105,200)(25,180,90) \ArrowArcn(105,200)(25,90,0) \ArrowArcn(105,200)(25,0,180)
\Photon(130,200)(210,200){3}{5} \Text(105,226)[]{$\times$} \Text(225,200)[]{+}

\Photon(240,200)(320,200){3}{5} \ArrowArcn(345,200)(25,180,0) \ArrowArcn(345,200)(25,0,270) \ArrowArcn(345,200)(25,270,180)
\Text(345,175)[]{$\times$} \Photon(370,200)(450,200){3}{5} \Text(470,200)[]{+}

\Photon(0,100)(80,100){3}{5} \ArrowArcn(105,100)(25,180,90) \ArrowArcn(105,100)(25,90,0) \ArrowArcn(105,100)(25,0,270) \ArrowArcn(105,100)(25,270,180) \Text(105,125)[]{$\times$} \Photon(130,100)(210,100){3}{5} \Text(105,75)[]{$\times$}   \Text(225,100)[]{+}

\Photon(240,100)(320,100){3}{5} \ArrowArcn(345,100)(25,180,120) \ArrowArcn(345,100)(25,120,60) \ArrowArcn(345,100)(25,60,0) \ArrowArcn(345,100)(25,0,180)  \Photon(370,100)(450,100){3}{5}  \Text(470,100)[]{+}  \Text(335,122)[]{$\times$} \Text(357.8,122)[]{$\times$}         

\Photon(0,0)(80,0){3}{5} \ArrowArcn(105,0)(25,180,120) \ArrowArcn(105,0)(25,120,60) \ArrowArcn(105,0)(25,60,0) \ArrowArcn(105,0)(25,0,180)  \Photon(130,0)(210,0){3}{5}  \Text(245,0)[]{+\hspace{2eM}$\cdots$}  \Text(95,-22)[]{$\times$} \Text(117.8,-22)[]{$\times$}
\vspace*{1cm}
\end{picture}
\vspace*{0.25cm}
\caption{Feynman diagrams in one loop approximation used to calculate the action (15).}
\end{figure}
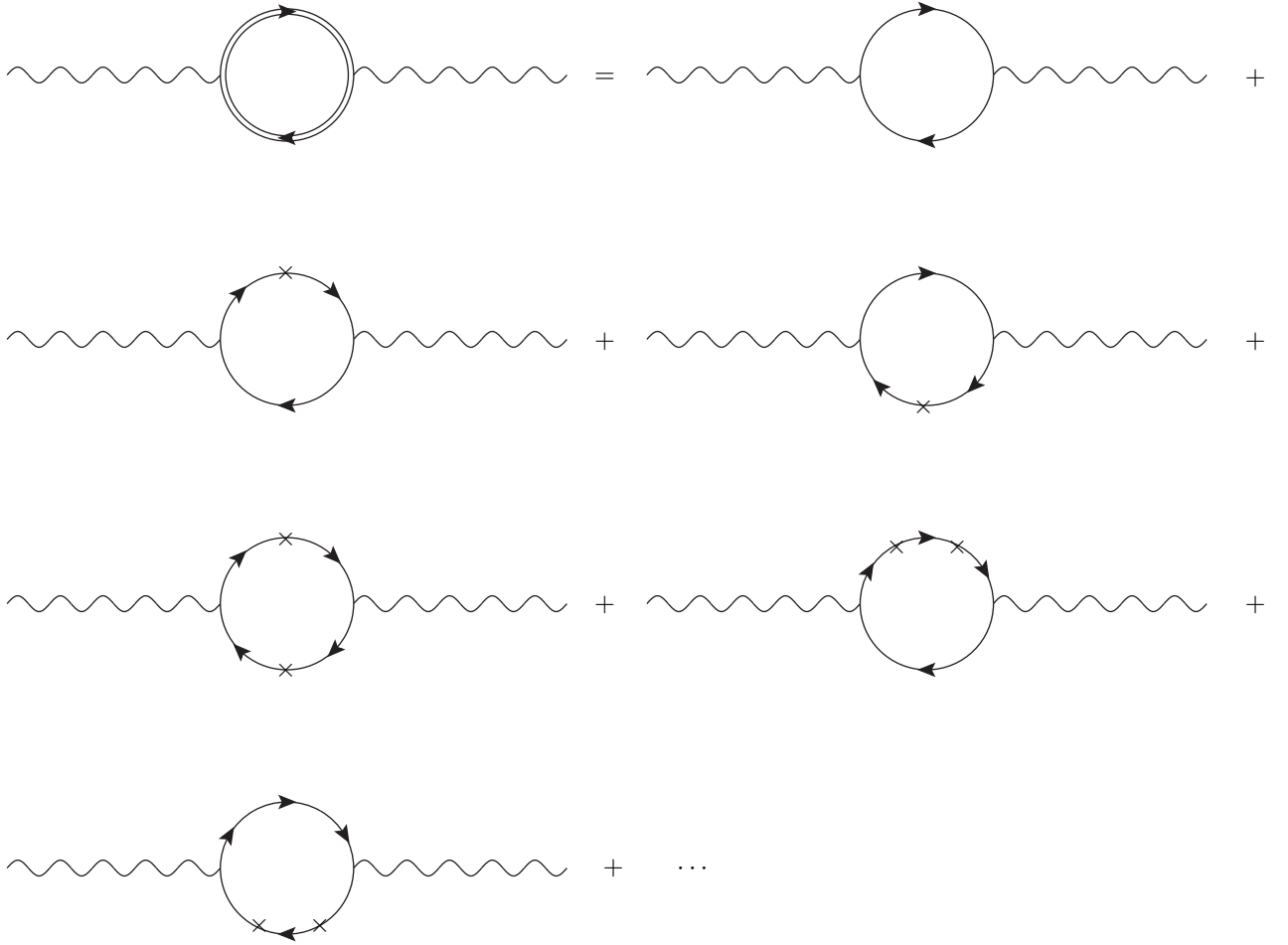

The radiative corrections will be obtained by introducing the expanded
fermion propagator (4). Thus, it follows that
\begin{multline}
 S_b(p)=\frac{i(\not\!p+m)}{p^2-m^2}+\frac{i}{(p^2-m^2)^2}(\not\!p+m) \not\!b\gamma_5  (\not\!p+m)\\
 +\frac{1}{(p^2-m^2)^3}(\not\!p+m) \not\!b\gamma_5  (\not\!p+m) \not\!b\gamma_5 (\not\!p+m) +\cdots\hspace{2cm}
\end{multline}

Using the notations

\begin{eqnarray}
P&=&\not\!p+m\\
B_5&=&\not\!b\gamma_5\\
{\cal D}^2&=&p^2-m^2
\end{eqnarray}
the propagator is written in compact form to linear terms in $B_5$:
\begin{equation}
 S_b(p)=\frac{i}{{\cal D}^2}P+\frac{i}{{\cal D}^4}PB_5P+\cdots
\end{equation}

Now, expanding the propagator $S_b (p-i\partial)$ to linear terms of
$\tilde{B}_5=\not\!b\gamma_5+i\!{\not\!\partial}$, it yields

\begin{equation}
 S_b(p-i\partial)=\frac{i}{{\cal D}^2}P+\frac{i}{{\cal D}^4}P\tilde{B}_5P+\cdots
\end{equation}
and

\begin{multline}
 S_b(p)\not\!\!AS_b(p-i\partial)\not\!\!A=-\frac{1}{{\cal D}^4}P\not\!\!AP\not\!\!A-\frac{1}{{\cal D}^6}P\not\!\!AP\tilde{B}_5P\not\!\!A\\
-\frac{1}{{\cal D}^6}PB_5P\not\!\!AP\not\!\!A-\frac{1}{{\cal D}^8}PB_5P\not\!\!AP\tilde{B}_5P\not\!\!A\,. \hspace{2.2cm}
\end{multline}

The terms that produce the CS induction are those that depend only on
a derivative of the field $A_\mu$:
$\not\!\!b\not\!\!\!A\not\!\!\partial\not\!\!A\gamma_5$ and that will
give the term structure of the CS-like term. This contribution comes
from the last term of the above expression. Thus, using the explicit
forms (17-18), it follows that

\begin{multline}
 PB_5P\not\!\!AP\tilde{B}_5P\not\!\!A=i\not\!p\not\!b\not\!p\not\!\!A\not\!p\not\!\partial\not\!p\not\!\!A\gamma_5+im^2\not\!p\not\!b\not\!p\not\!\!A\not\!\partial\not\!\!A\gamma_5+im^2\not\!p\not\!b\not\!\!A\not\!p\not\!\partial\not\!\!A\gamma_5\\
\\
+im^2\not\!p\not\!b\not\!\!A\not\!\partial\not\!p\not\!\!A\gamma_5-im^2\not\!b\not\!p\not\!\!A\not\!p\not\!\partial\not\!\!A\gamma_5-im^2\not\!b\not\!p\not\!\!A\not\!\partial\not\!p\not\!\!A\gamma_5\\
\\
-im^2\not\!b\not\!\!A\not\!p\not\!\partial\not\!p\not\!\!A\gamma_5-im^4\not\!b\not\!\!A\not\!\partial\not\!\!A\gamma_5+\cdots\hspace{1.2cm}
\end{multline}

In the above expression, the terms that do not contain derivatives and
those that are in odd number of Dirac matrices have been omitted since
their traces vanish. Now, the terms of the above expression of eight
and six to four $\gamma$ matrices are reduced using the properties
$\not\!c\not\!d=-\not\!d\not\!c+2(c\cdot d)$ e $\not\!c^2=c^2$, valid
for any four-vector $c^\mu$:

\begin{eqnarray}
 && ip^4\not\!b\not\!\!A\not\!\partial\not\!\!A\gamma_5  -2ip^2\not\!b\not\!\!A(p\cdot\partial)\not\!p\not\!\!A\gamma_5 -2ip^2(b\cdot p)\not\!p\not\!\!A\not\!\partial\not\!\!A\gamma_5 \nonumber\\
&&\nonumber\\
&&+4i(b\cdot p)\not\!p\not\!\!A(p\cdot\partial)\not\!p\not\!\!A\gamma_5+2im^2(b\cdot p)\not\!p\not\!\!A\not\!\partial\not\!\!A\gamma_5+2im^2\not\!p\not\!b\not\!\!A(p\cdot\partial)\not\!\!A\gamma_5\nonumber\\
&&\nonumber\\
&&-2im^2(p\cdot A)\not\!b\not\!\partial\not\!p\not\!\!A\gamma_5-im^4\not\!b\not\!\!A\not\!\partial\not\!\!A\gamma_5
\end{eqnarray}

The next step is to insert the above result in action (15) to
calculate the integrals in moments. It should be noted that, by
counting the powers, the integrals proportional to $p^4$ are
logarithmic divergence, while proportional to $p^2$ are finite:

\begin{eqnarray*}
 \int_{inf} \frac{d^4p}{(2\pi)^4}\frac{p^4}{(p^2-m^2)^4}\hspace{0.8cm}\mbox{,}\hspace{0.8cm} \int_{fin} \frac{d^4p}{(2\pi)^4}\frac{p^2}{(p^2-m^2)^4}\hspace{0.8cm}\mbox{e}\hspace{0.8cm}\int_{fin} \frac{d^4p}{(2\pi)^4}\frac{1}{(p^2-m^2)^4}\,.
\end{eqnarray*}

In an arbitrary dimension $D$, these finite demensionally regularized
integrals \cite{Peskin1995} are given by

\begin{eqnarray}
&& \int \frac{d^Dp}{(2\pi)^D}\frac{1}{(p^2-m^2)^\alpha}=\frac{(-1)^\alpha i}{(4\pi)^{D/2}m^{2\alpha-D}}\frac{\Gamma(\alpha-D/2)}{\Gamma(\alpha)}    \\
&&\nonumber\\
&&\int \frac{d^Dp}{(2\pi)^D}\frac{p_\mu p_\nu}{(p^2-m^2)^\alpha} = \frac{g_{\mu\nu}}{2}\frac{(-1)^{\alpha-1}i}{(4\pi)^{D/2}m^{2\alpha-D-2}}\frac{\Gamma(\alpha-D/2-1)}{\Gamma(\alpha)}
\end{eqnarray}

By explicitly computing the fifth term (finite) of the above
expression using the trace of the Dirac matrices,
$tr_D[\gamma^\mu\gamma^\nu\gamma^\rho\gamma^\sigma]=4i\varepsilon^{\mu\nu\rho\sigma}$,
and the integral (26) for $D=4$, it is obtained:

\begin{equation}
 -\frac{ie^2}{2}tr_D\int d^4x\int\frac{d^4p}{(2\pi)^4}\frac{2im^2(b\cdot p)\not\!p\not\!\!A\not\!\partial\not\!\!A\gamma_5}{(p^2-m^2)^4}=\frac{e^2}{48\pi^2}\int d^4x\,\varepsilon^{\mu\nu\rho\sigma}b_\mu A_\nu\partial_\rho A_\sigma
\end{equation}

The first four terms of expression (24) generate infinite terms in the
effective action. To perform the calculations, the regularization $D =
4-2\epsilon$ for the calculation of divergent integrals is
undertaken. While $\epsilon$ is kept not null, these originally
divergent integrals are kept finite and thus they can be either added
or removed. \par

This infinity integral is represented by
\cite{Peskin1995,Oliveira2010}

\begin{eqnarray}
 \int \frac{d^Dp}{(2\pi)^D}\frac{p_\mu p_\nu p_\rho p_\sigma}{(p^2-m^2)^4}=\frac{i}{384\pi^2}\left[\frac{1}{\epsilon}+\log\frac{4\pi}{m^2}-\gamma+{\cal O}(\epsilon)   \right](g_{\mu\nu}g_{\rho\sigma}+g_{\mu\rho}g_{\nu\sigma}+g_{\mu\sigma}g_{\nu\rho})\nonumber\\
\end{eqnarray}

The sum of infinity parts vanishes
\begin{eqnarray}
&&2e^2\int d^4x\,\varepsilon^{\mu\nu\rho\sigma}\left\{i.\frac{i}{16\pi^2}\left[\frac{1}{\epsilon}+\log\frac{4\pi}{m^2}-\gamma+{\cal O}(\epsilon)  \right]b_\mu A_\nu\partial_\rho A_\sigma\right.\nonumber\\
&& \left. \right.\nonumber\\
&&\left. -2i.\frac{i}{64\pi^2}\left[\frac{1}{\epsilon}+\log\frac{4\pi}{m^2}-\gamma+{\cal O}(\epsilon)  \right]b_\mu A_\nu\partial_\rho A_\sigma \right.\nonumber\\
&& \left. \right.\nonumber\\
 && \left.-2i.\frac{i}{64\pi^2}\left[\frac{1}{\epsilon}+\log\frac{4\pi}{m^2}-\gamma+{\cal O}(\epsilon)  \right]b_\mu A_\nu\partial_\rho A_\sigma \right.\nonumber\\
&& \left. \right.\nonumber\\ 
&&\left.+4i.\frac{i}{384\pi^2}\left[\frac{1}{\epsilon}+\log\frac{4\pi}{m^2}-\gamma+{\cal O}(\epsilon)  \right](b_\mu A_\nu\partial_\rho A_\sigma -b_\mu A_\nu\partial_\rho A_\sigma)    \right\}=0\nonumber
\end{eqnarray}

So, the induced term depends only on the finite part of the effective action and is given by
\begin{equation}
 S_{CS}^{(3+1)D}=\frac{e^2}{12\pi^2}\int d^4x\varepsilon^{\mu\nu\rho\sigma}b_\mu A_\nu\partial_\rho A_\sigma\,.
\end{equation}

from which the following relation is obtained
\begin{equation}
 \eta_\mu=\frac{e^2}{6\pi^2}b_\mu\,.
\end{equation}

Therefore, it is concluded that the addition of a term with a
background field that breaks the Lorentz symmetry of the Lagrangian
gauge theory leads to the usual Chern-Simons type action in the
four-dimensional spacetime. As is well noted in the literature, this
term is finite and obtained by various regularization methods. The
only difference is the constant of proportionality, which depends
exclusively on the type of the regularization method used for any
specific calculations.

\section{\label{sec:IV}Conclusions}

In this work, the induced Chern-Simons like action in a
quadrimensional space was calculated. It should be noted that this
accomplishment is possible even when starting from a Lagrangian that
does not contain this specific term. That is only possible if we the
term of the Lorentz symmetry breaking is introduced in the QED
lagrangian, since the term $-\!\not\!b\gamma_5$ induces the
calculation of $tr_D[\gamma^\mu\gamma^\nu\gamma^\rho\gamma^\sigma]$,
which generates the necessary structure
$\varepsilon^{\mu\nu\rho\sigma}$ relevant to that induced term. The
procedure for obtaining such action had as its starting point the
expansion of the modified fermion propagator given by the new
theory. This result is in agreement with those obtained in the
literature, which may be distinct in regards to a constant if
different methods of regularization are used (see, e.g., \cite{Ebert2004,Alfaro2006,Mariz2006,Nascimento2007,Andrianov2008,Santos2009}). In the present
calculation, using the method of dimensional regularization, it is
noted that the induced action of the different terms cancel one another, and the limit $m\to0$ is not necessary and the induced action is finite.

\vspace{1cm}

\textbf{Acknowledgments. }The author is grateful to Professor Adilson José da Silva for discussions on the subject and to CNPq (Conselho Nacional de Deselvolvimento Científico e Tecnológico) for financial support.

\end{document}